\documentstyle[aps,prl]{revtex}
\begin{document}
\input epsf.sty
\twocolumn[\hsize\textwidth\columnwidth\hsize\csname %
@twocolumnfalse\endcsname
\draft
\widetext
%%%%%%%% prl (above) %%%%%%%%%%%%%%%%%%

\title{Reentrant Spin-Peierls Transition in Mg-Doped CuGeO$_3$}

\author{V. Kiryukhin, Y. J. Wang, S. C. LaMarra, R. J. Birgeneau}
\address{Department of Physics and Center for Materials Science 
and Engineering,
Massachusetts Institute of Technology,
Cambridge, Massachusetts 02139}
\author{T. Masuda, I. Tsukada, K. Uchinokura}
\address{Department of Applied Physics, The University of Tokyo,
7-3-1 Hongo, Bunkyo-ku, Tokyo 113-8656, Japan}

\date{\today}
\maketitle

\begin{abstract}

We report a synchrotron x-ray scattering study of 
the diluted spin-Peierls (SP) material Cu$_{1-x}$Mg$_x$GeO$_3$. 
In a recent paper we have shown that the SP dimerization attains
long-range order only for $x<x_c\sim$0.022$\pm$0.001. Here we report that
the SP transition is reentrant in the vicinity of the 
critical concentration $x_c$. This is manifested by  
broadening of the SP dimerization superlattice peaks below the reentrance
temperature, T$_r$, which may mean either the complete loss of the 
long-range SP order or the development of a short-range ordered component
within the long-range ordered SP state. 
Marked hysteresis and very large relaxation times are found in the
samples with
Mg concentrations in the vicinity of $x_c$. The reentrant SP transition 
is likely related to the competing N\'eel transition which occurs at a
temperature similar to T$_r$. We argue that 
impurity-induced competing interchain interactions
play an essential role in these phenomena.

\end{abstract}


%%%%%% prl format (below) %%%%%%%%%%%%%%
\phantom{.}
]
\narrowtext
%%%%%%%%% prl (above) %%%%%%%%%%%%%%%%%%

Low-dimensional quantum spin systems exhibit a variety of intriguing  
and often counter-intuitive properties.
A prominent example of such a material is the spin-Peierls (SP) system \cite{SP}
which consists of an array of one-dimensional (1D) antiferromagnetic
spin-chains with S=$1\over 2$ on a deformable 3D lattice. Below the
spin-Peierls transition temperature, T$_{SP}$, the spin-chains dimerize
and a gap opens in the magnetic excitation spectrum. The discovery of an 
inorganic SP compound CuGeO$_3$ (Ref. \cite{Hase})
made possible a systematic study of impurity
effects on SP systems \cite{Hase2}. In CuGeO$_3$, 
Zn$^{2+}$, Mg$^{2+}$ (S=0), and Ni$^{2+}$ (S=1) 
can be readily substituted for Cu$^{2+}$ (S=$1\over 2$)
thus directly affecting the
spin-chains \cite{Luss,Masuda}. 
Generally, it is found that with increasing impurity concentration,
the SP transition temperature rapidly decreases and a N\'eel state
appears at low temperatures \cite{Luss,Masuda}.

While a large amount of work has been devoted to the doped CuGeO$_3$
system since its discovery, the
temperature-doping (T-$x$)
phase diagram of this system has recently been substantially
revised \cite{Masuda}-\cite{Masuda2}. 
In particular, Masuda {\it et al.} \cite{Masuda,Masuda2} have shown that
in the Mg-doped compound, the SP transition abruptly disappears 
at a critical Mg concentration $x_c\sim 0.023$, 
and the N\'eel temperature T$_N$
exhibits a conspicuous jump at the same Mg concentration. 
The staggered magnetic moment in the N\'eel phase and the magnitude of the
SP lattice displacement were also found to change substantially 
in the vicinity of $x_c$
\cite{Nakao}. Moreover, in the vicinity of $x_c$, two peaks in the
temperature dependence of the magnetic susceptibility were found \cite{Masuda2};
these peaks were attributed to two separate N\'eel transitions, and 
coexistence of two different phases (phase separation)
for $x\sim x_c$ was proposed.
Zn-doped CuGeO$_3$ exhibits very similar properties, and therefore this
behavior appears to be universal for diluted CuGeO$_3$ in which the Cu$^{2+}$
is replaced by a non-magnetic ion.

We have recently reported high resolution synchrotron x-ray diffraction 
measurements on high-quality single crystals of Mg-doped CuGeO$_3$ \cite{Wang}. 
We found that while measurable
SP lattice dimerization persists for $x$ larger than $x_c$, 
the SP dimerized state attains long range order (LRO) only for $x<x_c$.
Moreover, for $x$ in the vicinity of $x_c$, SP LRO
is achieved at a temperature which is significantly lower than the SP
transition temperature determined in magnetic susceptibility or heat
capacity measurements \cite{Wang}. 
We have proposed that these unusual phenomena 
result from competing interactions that are inevitably present in
any diluted SP material. These results clearly demonstrate that  
synchrotron x-ray diffraction is an extremely valuable experimental 
method to study doped SP materials. However, the important
low-temperature region
of the T-$x$ phase diagram which contains the N\'eel phase has thus far 
not been thoroughly investigated with this experimental technique.

In this paper, we report a synchrotron x-ray scattering study of the
T-$x$ phase diagram of Cu$_{1-x}$Mg$_x$GeO$_3$ with 
emphasis on the low-temperature regime. We find that
for the Mg concentration $x$ in the vicinity of $x_c$ the SP transition
is reentrant.  This is manifested by
broadening of the SP dimerization superlattice peaks below the reentrance
temperature, T$_r$. This broadening most likely results from the loss
of SP long-range order below T$_r$ for samples with $x\lesssim x_c$. 
However, we cannot exclude
another possible scenario in which a short-range ordered (SRO) component
with significant volume fraction
develops within the long-range ordered SP state below T$_r$. In this
case, a two-phase state is realized. 
Marked hysteresis and very large relaxation times are found in the
samples with
Mg concentrations in the vicinity of $x_c$. The reentrant SP transition
may be related to the competing N\'eel transition which occurs at the
temperature similar to T$_r$. As we argued in Ref. \cite{Wang}, we believe
that the structural and related magnetic properties of doped CuGeO$_3$
are strongly influenced by impurity-induced competing interchain
interactions and, possibly, random fields, and therefore are
similar to the properties of other systems with competing interactions
and/or fields, such
as Spin Glasses (SG) and Random Field Ising Model (RFIM) compounds.
Here we propose that the reentrant SP state can be naturally explained 
by the effects of the competing interactions since such a
reentrant transition is commonly found in SG compounds.  
 
The experiment was carried out at MIT-IBM beamline X20A at the 
National Synchrotron Light Source at Brookhaven National Laboratory. 
The 8.5 keV 
x-ray beam was focused by a mirror,
monochromatized by a pair of Ge (111) crystals, scattered from the sample,
and analyzed by a Si (111) analyzer.   
High quality Cu$_{1-x}$Mg$_x$GeO$_3$ single crystals with $x$=0.017, 0.0209,
and 0.0229 from the same batches as those studied in Ref. \cite{Masuda}
were used. Carefully cleaved samples were loaded into a helium flow cryostat;
the lowest temperature achievable in this setup was 1.4 K. The experiment was 
carried out in the vicinity of the (1.5, 1, 1.5) SP dimerization peak 
position with the (H K H) zone in the horizontal scattering plane. 
Longitudinal
(parallel to the scattering vector) and transverse scans in the (H K H)
zone were collected. In the direction
perpendicular to the scattering plane, the intensity was 
automatically integrated due to our experimental setup.
Since the beamline was optimized for the vertical scattering geometry
rather than the horizontal scattering geometry required by the helium cryostat,
the spatial resolution of this experimental setup was worse than that 
achieved in our previous work, Ref. \cite{Wang}. To determine the resolution,
we used our previous result that at temperatures of order 5--6 K the
$x$=0.017 and $x$=0.0209 samples attain
long range order, which was defined as a state with a correlation
length $\xi$ of at least five thousand $\rm\AA$ngstroms \cite{Wang};
significantly larger correlation lengths are essentially macroscopic.
To extract the intrinsic correlation
length, the data for all samples were fitted to a convolution of the
measured resolution function with the intrinsic cross section. 
Several different intrinsic line-shapes produced fits of similar
quality. To be consistent with the data analysis of Ref. \cite{Wang},
we use 3D Lorentzian-squared line-shapes in this paper.

%%==============================================================================
\begin{figure}
\centerline{\epsfxsize=2.9in\epsfbox{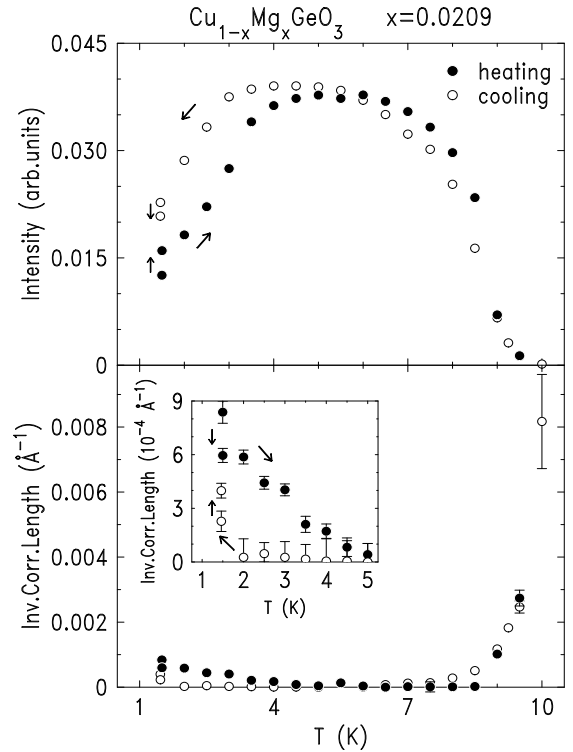}}
\vskip 5mm
\caption{Temperature dependencies of the (1.5, 1, 1.5) SP peak intensity
(top panel) and the corresponding longitudinal inverse correlation length
(bottom panel) taken on heating and on cooling in the x=0.0209 sample.
The inset shows the low-temperature inverse correlation length.}
\label{fig1}
\end{figure}
%%========================================================================

We first consider the most interesting case when the impurity concentration
$x$ is very close to but slightly smaller than the critical concentration $x_c$.
Fig. 1 shows the temperature dependencies of the (1.5, 1, 1.5) SP dimerization
peak intensity and the corresponding longitudinal
inverse correlation length taken on
heating and on cooling in the $x$=0.0209 sample. 
The N\'eel temperature in this sample is T$_N\sim$2.7 K \cite{Masuda2};
the decrease of the SP peak intensity at low temperatures 
appears to be initiated at the N\'eel transition.
The data were taken
as follows: first, the sample was quickly cooled down to T=1.4 K (15 seconds 
from T$>$10 K to T$<$2 K), then the data were taken on slow heating and
on slow cooling, with each run lasting approximately
12 hours. According to our recent
high-resolution measurements \cite{Wang}, this sample is in the LRO SP state
at T$\sim$4 K. However, as the inset in Fig. 1 shows, the long-range order
is lost at lower temperatures. Therefore, the SP transition in this sample
is reentrant. 

In Fig. 2 we show transverse scans at the (1.5, 1, 1.5) SP dimerization
peak position taken at T=5 K and T=1.4 K in the cooling run. The peak 
intensities were scaled to be the same. The broadening of the SP peak at
T=1.4 K is small but definitely non-zero. We do not believe that this
broadening is an experimental artifact arising, for example, from an anomaly
in the background.
The solid lines are the results of fits to the convolution of the measured
experimental resolution function with the intrinsic 3D Lorentzian-squared
line shape, which in our case reduces to a Lorentzian to the 1.5 power
scattering profile
for the in-plane scans due to the out-of-plane integration.   
The inverse correlation
length $\xi^{-1}$ at T=1.4 K determined in such a way from both the 
longitudinal and transverse scan is 2.5$\times 10^{-4}\rm\AA^{-1}$.
The exact value of the low-temperature correlation length should be
taken with caution since fits to different intrinsic line shapes produced
fits of similar quality with correlation length values deviating from the
above result by as much as a factor of 2. Due to this ambiguity in the
line-shape analysis, we also cannot exclude the 
possibility that the peak broadening is due to the development of a 
large volume short-range ordered component in the otherwise LRO system.
This does not affect, however, the important qualitative result 
that the SP transition is reentrant at low temperature in this sample. 

%%==============================================================================
\begin{figure}
\centerline{\epsfxsize=2.7in\epsfbox{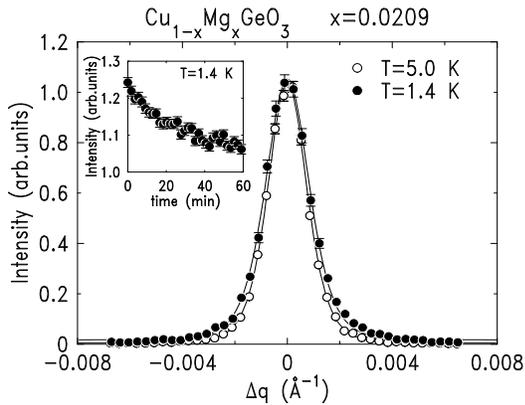}}
\vskip 5mm
\caption{Transverse scans at the (1.5, 1, 1.5) SP peak position in the 
$x$=0.0209 sample at T=5 K and T=1.4 K taken on cooling. 
The peak intensities were scaled to be the same. The solid lines are the
results of fits as discussed in the text. The inset shows the time dependence 
of the SP peak intensity after the sample was cooled from T=2 K to T=1.4 K.}
\label{fig2}
\end{figure}
%%=====================================================================

The data of Fig. 1 show marked hysteresis. 
The hysteresis results from anomalously
slow dynamics in the samples with $x$ in the vicinity of $x_c$. The inset
in Fig. 2 illustrates this phenomenon. In the inset we
show the time dependence of the SP
peak intensity after the sample was cooled from T=2 K to T=1.4 K. Apparently,
1 hour is not sufficient for the sample to come to equilibrium. The slow 
dynamics at T=1.4 K in this sample is also illustrated in Fig. 1: the 
two different data points for each run represent data taken shortly
after cooling and after an additional 5 hours at T=1.4 K. The direction of
change is shown with arrows. The anomalously slow dynamics is most pronounced
in the regions of the largest hysteresis, that is, near the SP transition 
temperature and below the N\'eel temperature. 

Samples with the Mg concentration $x$ slightly larger than $x_c$ exhibit
behavior very similar to the behavior of the $x$=0.0209 sample with the
exception that the SP peaks are broadened at all temperatures. Fig. 3
shows the temperature dependencies of the SP peak intensity and the
corresponding longitudinal inverse correlation length for the
$x$=0.0229 sample taken on heating and on cooling. The heating-cooling
protocol was the same as that for the $x$=0.0209 sample. Very large
suppression of the SP peak intensity below T$_N$, extra broadening of the
SP peak at low temperatures, marked hysteresis, and anomalously slow dynamics
are found in this sample, similar to the results in the $x$=0.0209 sample. 
A similar broadening of the SP reflections and suppression of the SP peak 
intensity in the {\it short-range} ordered state has been recently reported
by Nakao {\it et al.} \cite{Nakao} 

%%==============================================================================
\begin{figure}
\centerline{\epsfxsize=2.7in\epsfbox{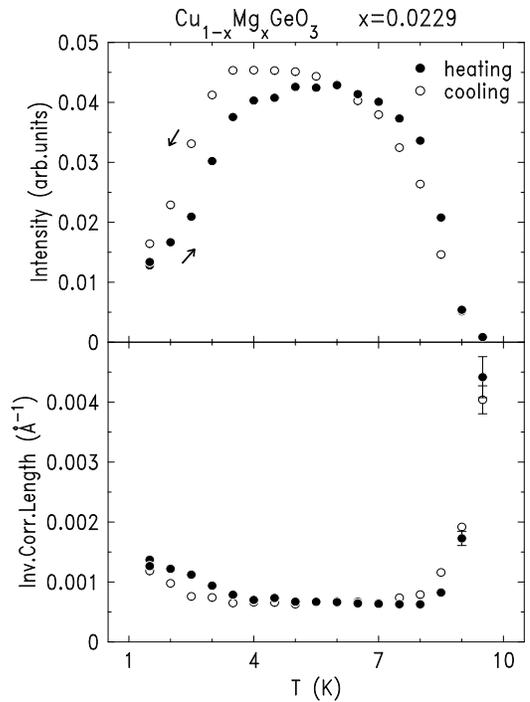}}
\vskip 5mm
\caption{Temperature dependencies of the (1.5, 1, 1.5) SP peak intensity
and the corresponding inverse correlation length in the $x$=0.0229 sample on
heating and on cooling.}
\label{fig3}
\end{figure}
%%=====================================================================

Samples with $x$ much lower than $x_c$ show different behavior. Fig. 4
shows the temperature dependencies of the SP peak intensity and width
for the $x$=0.017 sample. The N\'eel temperature in this sample 
is T$_N\sim$2.1$\pm$0.1 K (Ref. \cite{Masuda2}). 
The decrease of the SP peak intensity
below T$_N$ is very small, and the hysteresis is much less pronounced
than in the case of $x\sim x_c$. The hysteresis and the accompanying
slow dynamics are also virtually absent for samples of 
Cu$_{1-x}$Mg$_x$GeO$_3$ with Mg concentrations much 
larger than $x_c$. 

As we noted in Ref. \cite{Wang}, we cannot rigorously exclude the scenario
in which for $x<x_c$ the low-temperature SP correlation length saturates
at some finite value that is larger than our resolution limit. Therefore,
it is possible that true LRO is never achieved in the $x$=0.0209 sample.
However, if a true phase boundary between the LRO and the SRO SP regions
does exist in the T-$x$ phase diagram of Cu$_{1-x}$Mg$_x$GeO$_3$, the 
impurity concentration $x$=0.0209 is very close to the true critical
concentration $x_c$ because of the large lower limit that Ref. \cite{Wang}
puts on the SP correlation length at T=4 K. The decrease of the correlation
length at low temperature in this sample, therefore, is almost certainly
associated with the reentrant character of the true phase boundary.
Another possible scenario \cite{Masuda,Masuda2} is that a first order 
transition as a function of $x$ occurs at $x$=$x_c$. In this case, 
two-phase coexistence (phase separation) takes place in a non-zero region
around $x_c$, and the impurity concentration $x$ at which SP LRO is
established in the entire sample
might not be precisely defined. In this scenario, $x$=0.0209 is
smaller that $x_c$ \cite{Masuda2}, and the low-temperature decrease of the
correlation length is again associated with the reentrant phase boundary
of the SP phase. 

The rich and complex properties of diluted CuGeO$_3$ described here and 
also in Refs. \cite{Wang,Masuda2} clearly 
cannot be explained by simple dilution
effects alone. We believe, however, that these properties can be
consistently explained by taking into account the close intrinsic analogy
between doped SP compounds
and other disordered systems with competing fields or
interactions, such as the RFIM compounds \cite{Bob} and spin glasses
\cite{Maletta}. In Ref. \cite{Wang}, we have argued that in-chain dilution
of a SP compound induces competing interactions and, concomitantly,
frustration in the dimerized system. Briefly, it is energetically favorable
to change the phase of the dimerization across the impurity in an isolated
chain, because otherwise an unpaired spin is created. 
The (mean field) interaction with the neighboring chains is,
on the other hand, minimized when the dimerization phase is constant, thus
creating competing interactions and frustration.           
Since the system can be mapped onto an effective 3D Ising model in which
the two dimer configurations possible in a given chain are associated with 
the up and down states of the Ising spins,
this is analogous to the situation in 3D Ising systems with mixed ferromagnetic
and antiferromagnetic bonds \cite{Wang}.
Therefore, it is natural to expect that
doped SP materials share many common properties with
spin glass systems. There also are higher order random field effects.
The properties of the latter systems 
are extensively discussed in the literature \cite{Bob,Maletta}, so that the
above analogy can significantly improve our understanding of doped
SP materials.

%%==============================================================================
\begin{figure}
\centerline{\epsfxsize=2.8in\epsfbox{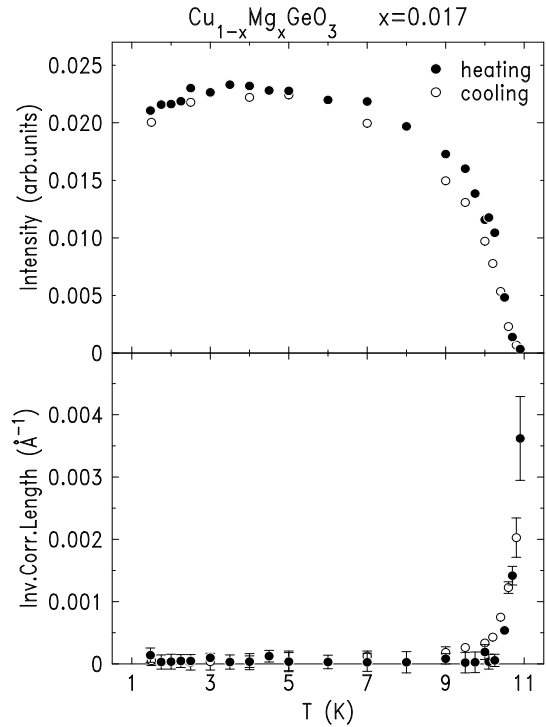}}
\vskip 5mm
\caption{Temperature dependencies of the (1.5, 1, 1.5) SP peak intensity
and the corresponding inverse correlation length in the $x$=0.017 sample on
heating and on cooling.} 
\label{fig4}
\end{figure}
%%==================================================================

The mapping into Ising magnet with random ferromagnetic and antiferromagnetic 
bonds is especially useful for understanding of the reentrant 
behavior and the slow dynamics found in Cu$_{1-x}$Mg$_x$GeO$_3$ because 
these phenomena are commonly found in SG compounds \cite{Maletta,Aeppli}. 
In the heuristic model proposed by Aeppli {\it et al.} \cite{Aeppli}, 
the destruction of the LRO at the reentrant transition 
in SG materials is attributed to the effects of random fields imposed on the
LRO network by randomly frozen finite spin clusters that form in these
materials at low temperatures. Expressed in the Ising pseudospin language,
a similar scenario is likely valid for doped SP compounds. Additional 
evidence for the important role of the induced random-field effects in 
Cu$_{1-x}$Mg$_x$GeO$_3$ comes from the anomalous behavior exhibited by
this compound in the vicinity of the high-temperature SP transition 
\cite{Wang}. 

Evidently, the N\'eel and the SP order parameters in doped CuGeO$_3$ are 
coupled to each other. At low dilution levels, 
both the SP dimerization and the N\'eel order coexist in a complex 
macroscopically uniform state in which both the SP and the AF correlations
retain LRO \cite{Fukuyama}.
This picture is consistent with our data for the $x$=0.017 sample. 
However, for higher impurity concentrations, the SP peak intensity is
dramatically suppressed at temperatures near 
T$_N$, indicating the competing character of the
SP and the N\'eel states. In addition, the structural
reentrance temperature T$_r$ and the magnetic ordering temperature
T$_N$ in the $x$=0.0209 sample are similar. Taking into account 
the competition between these two states and the
analogy to the SG reentrant behavior described above,
we propose that the main features of the
T-$x$ phase diagram of Cu$_{1-x}$Mg$_x$GeO$_3$ can be
explained as follows. Competing interactions and/or fields play an essential
role in this material. This accounts for the spin-glass-like behavior at the 
structural SP transition \cite{Wang}, and for the destruction of the SP LRO
at low temperatures for $x\sim x_c$. The self destruction of the SP state
results in the jump of the N\'eel temperature for concentrations around
$x=x_c$. It is also 
plausible that the reentrant character of the SP SRO state is one of the 
factors that define the N\'eel temperature for $x>x_c$, and that the 
double-peak structure of the magnetic susceptibility in the vicinity of $x_c$
\cite{Masuda2} is also related to this phenomenon. We should, however, note
that other descriptions \cite{Mostovoy}
of the phase behavior of Cu$_{1-x}$Mg$_x$GeO$_3$,
notably the existence of a tricritical point and the corresponding
first order transition line with its associated two-phase coexistence
(phase separation)
\cite{Masuda,Masuda2}, are possible.

In conclusion, we have carried out a synchrotron x-ray scattering study of the 
diluted SP material Cu$_{1-x}$Mg$_x$GeO$_3$ with emphasis on the 
behavior at low temperatures. We have found that
in the vicinity of $x_c$, the SP transition is reentrant and that the 
system exhibits anomalously slow dynamics and marked hysteresis
in the reentrant temperature regime.
We believe that these phenomena can be consistently described 
by a model that incorporates the close analogy between the doped 
SP system and other disordered systems with competing interactions/fields
such as SG and RFIM compounds. Clearly, further theoretical work 
along these lines is called for.

We are grateful to G. Shirane, J. P. Hill, and L. D. Gibbs for important
discussions. This work was supported by the NSF under Grant No. 
DMR97-04532. This work was also supported by NEDO International Joint
Research Grant, and by Grant-in-Aid for COE Research ``SCP Coupled System''
from the Ministry of Education, Science, Sports, and Culture of Japan, 
and by a Research Fellowship of the Japan Society for the 
Promotion of Science for
Young Scientists (T. M.).

%%%%%%%%%%%%%%%%%%%%%%%%%%%%%%%%%%%%%%%%%%%%%%%%%%%%%%%%%%%%%%%%%%%%%%%%

\end{document}